\documentclass[12pt]{iopart}
\usepackage{iopams} 
\usepackage{graphics}
\usepackage{pennames}
\usepackage{amssymb}
\usepackage{setstack}
\usepackage{color}

\def\mt{m_{\tt T}}

\newcommand{\RCP}{R_{\rm CP}}
\def\pt{p_{\tt T}}

\newcommand{\av}[1]{\left\langle #1 \right\rangle}
\newcommand{\gev}{\mathrm{GeV}}
\newcommand{\sqrtsNN}{\sqrt{s_{\scriptscriptstyle{{\rm NN}}}}}

\renewcommand{\d}{{\rm d}}

\begin{document}
\title[NA57 main results]{NA57 main results} 
\author{G E Bruno for
the NA57 Collaboration~\footnote[1]{For the full author list see
Appendix ``Collaborations'' in this volume.}}
\address{
Dipartimento~IA~di~Fisica~dell'Universit{\`a}~e~del~Politecnico~di~Bari~and~INFN,~Bari,~Italy}
\ead{Giuseppe.Bruno@ba.infn.it}
\begin{abstract}
The CERN NA57 experiment was designed to study the production of
strange and multi-strange particles in heavy ion collisions at SPS energies; 
its physics programme 
is essentially completed.  
A review of the main 
results is presented.
\end{abstract}
%
%
%
\section{Introduction} 
The NA57 experiment at CERN was proposed~\cite{proposal} to 
study (multi-)strange baryon and anti-baryon production  
in Pb--Pb collisions at SPS. 
Its principal aim was to 
refine  
and extend the measurements of its  
predecessor WA97, which first measured the enhancements of hyperons  
and anti-hyperons in Pb--Pb collisions with respect to reference pBe  
collisions~\cite{WA97Enh}.  
The observed pattern of enhancements increasing with the particle's  
strangeness content constituted one of the main pieces of evidence for 
the creation, in Pb-Pb collisions at the SPS, of a new state of matter 
displaying many of the predicted features of a deconfined system~\cite{CERN_ANN}.  

The main contributions of NA57 fall into three areas:  hyperon enhancements, strange
particle spectra, and nuclear modification factors.
\section{Experimental technique}
Strange particles were detected by reconstructing their 
weak decays into final state containing charged particles only, 
i.e., \PgL$\rightarrow$\Pp\Pgpm,
\PgXm$\rightarrow$\PgL\Pgpm$\rightarrow$\Pp\Pgpm\Pgpm,
\PgOm $\rightarrow$\PgL\PKm$\rightarrow$\Pp\Pgpm\PKm
(and c.c. for anti-hyperons), and \PKzS $\rightarrow$ \Pgpm\Pgpp.
The charged tracks emerging from strange particle decays were detected  
in a telescope made from an array of silicon pixel detectors of 5x5 cm$^2$\ cross-section
placed in an approximately uniform magnetic field;   
the bulk of the detectors was closely packed in approximately 30 cm length, this part being
used for pattern recognition. The telescope was placed above the beam line, inclined and
aligned with the lower edge of the detectors laying on a line pointing to the target.
The inclination angle and the distance   
from the target were set so as to accept particles produced
in about a unit of rapidity around mid-rapidity, with
transverse momentum above a few hundred MeV/$c$.  
The selected samples of strange and multi-strange particles after final 
kinematical cuts were  
almost background free ($noise/signal$\ at most a few percents).  

The centrality of the collisions was determined by analysing the charged particle
multiplicity measured by two stations of microstrip silicon detectors (MSD) which together sampled 
the pseudorapidity interval $2<\eta<4$. 
Detailed descriptions of the NA57 experimental apparatus, hyperon selection strategy and centrality 
determination can be found in 
references~\cite{enh160,Multiplicity}.

\section{Hyperon enhancements}
An enhanced production of strange particles in nucleus--nucleus collisions with respect  
to proton--induced reactions was suggested long time ago as a possible signature of the 
phase transition from colour-confined hadronic matter to a quark-gluon plasma (QGP)~\cite{rafelski}. 
Moreover, the magnitude of the enhancement was predicted to increase with the strangeness
content of the particles.  The argument being that the $s$\ and $\bar{s}$\ quarks, 
abundantly produced in the deconfined phase, would recombine to form strange and multi-strange 
particles in a time much shorter than that required to produce them by successive rescattering  
interactions in a hadronic gas.  

\begin{figure}[hb]
\centering
\resizebox{1.0\textwidth}{!}{%
\includegraphics{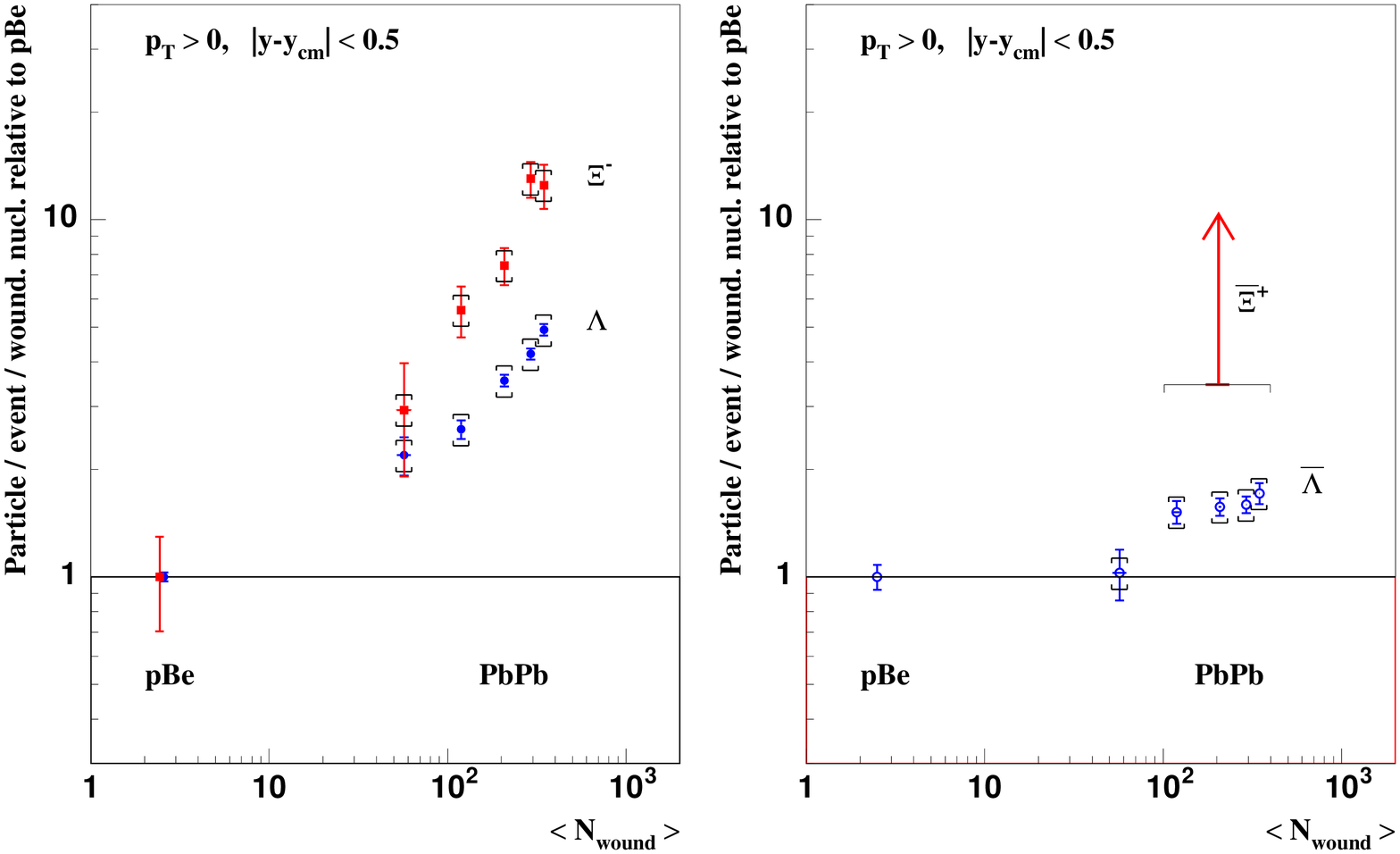}
\includegraphics{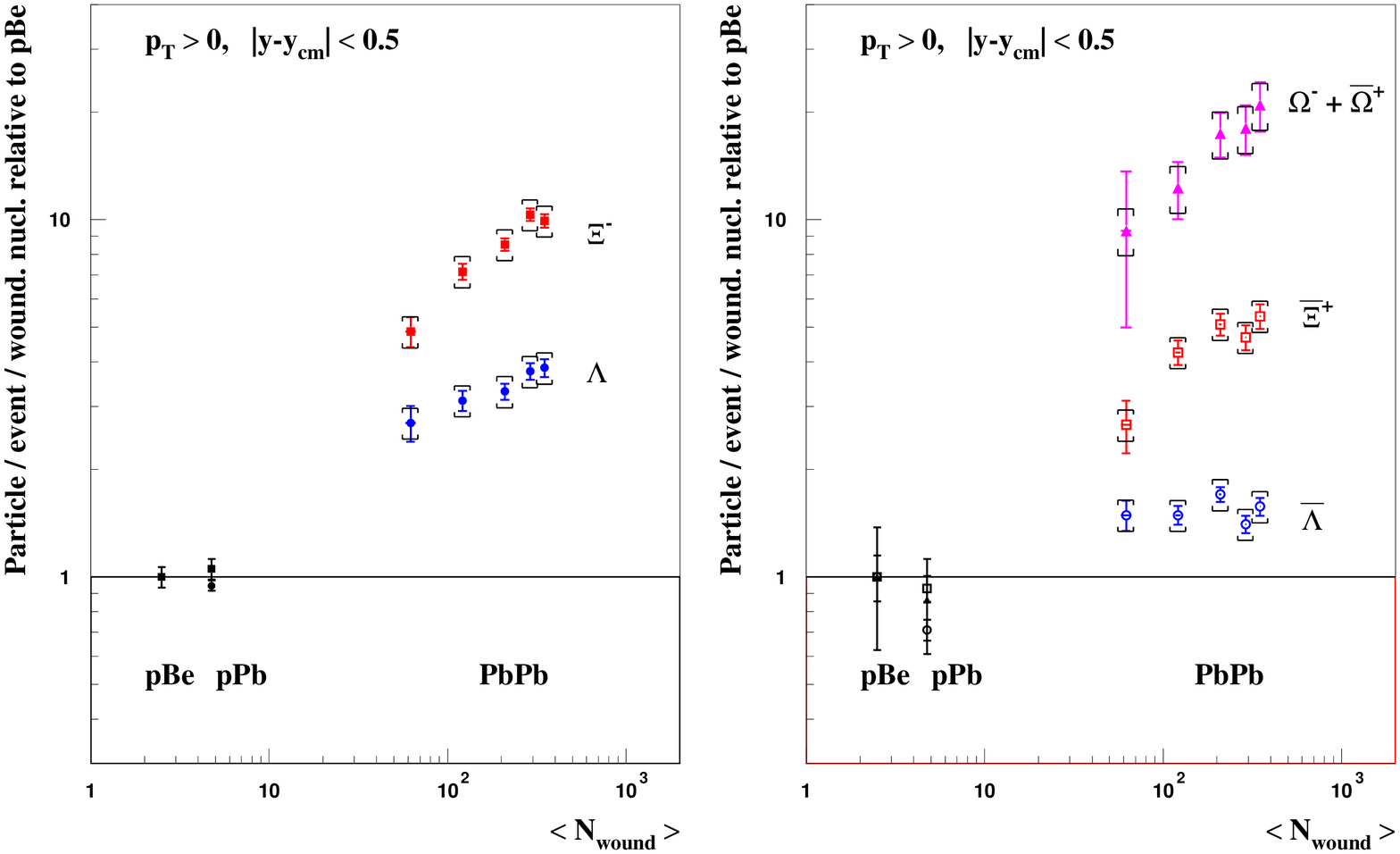}}
\caption{Hyperon enhancements as a function of the number of participants ($N_{wound}$)
at 40 (1$^{st}$\ and 2$^{nd}$ panels, ref~\cite{Prova1}) and 158  
(3$^{rd}$\ and 4$^{th}$\ panels, ref~\cite{enh160}) $A$\ GeV/$c$.
The symbol $_{\sqcup}^{\sqcap}$\  shows the systematic error, 
the arrow in the 2$^{nd}$ panel indicates the lower limit to the \PagXp\ 
enhancement in the four most central classes at 95\% confidence level.}
\label{fig:enh}
\end{figure}

The enhancements are defined as the central rapidity particle yield per participant 
($E=Y/<N_{wound}>$, $Y=\int^{y_{cm+0.5}}_{y_{cm-0.5}} \,
{\rm d}y\int^{\infty}_{0}\frac{{\rm d}N^2}{{\rm d}\pt {\rm d}y}\,{\rm d}\pt$) 
relative to pBe collisions.  
The hyperon enhancements measured by NA57 at 
40 and 158 $A$\ GeV/$c$\ are shown as a  
function of centrality in figure~\ref{fig:enh}.  

The results presented above confirm, refine and extend the study of strangeness  
enhancements initiated by the WA97 experiment~\cite{WA97Enh}.  
In particular  the increase of the magnitude of the enhancement with the strangeness
content of the particles is confirmed, as expected in a QGP scenario~\cite{rafelski}. 
The highest enhancement is measured for the triply-strange $\Omega$ hyperon
and amounts to about a factor 20 in the most central class. 
Conventional hadronic models fail to reproduce this result.    
The confirmation of the WA97 results is important in itself; however NA57 could 
measure the hyperon 
enhancement with greater detail, 
as a function of centrality, and of energy.  
\newline
{\bf Centrality dependence} 
A comprehensive discussion of the centrality dependence of the enhancements 
can be found in reference~\cite{enh160}.  
Here it is worth noting that (at 158 A GeV/$c$) the normalized yields 
for the pBe and pPb data are compatible with each other within the error limits, as expected  
from $N_{\rm wound}$\ scaling.  
In Pb--Pb collisions the hyperon production is already significantly enhanced at $N_{wound}$\ 
as low as about 50, and a significant centrality dependence of the 
enhancements for all particles except $\overline\Lambda$ is observed. 
For the two most central classes
($\simeq 10\%$ of most central collisions) a saturation of the enhancements
cannot be ruled out, in particular for \PgXm and \PagXp.  
In contrast, the flatness of the \PagL\ enhancement with centrality is somewhat puzzling.  
This could be due to some difference in the initial production mechanism, or it could be the consequence
of, e.g., a  centrality-dependent \PagL\ absorption in a nucleon-rich environment.  
It is worth recalling that (i) within the NA57 acceptance  
$\overline\Lambda$ is the only strange baryon for which we observe a significantly non-flat rapidity 
distribution~\cite{rapidity}, and  
(ii) a similar behaviour of \PagL\ enhancement with centrality at SPS top energy 
has been confirmed by NA49~\cite{BlumeSQM07}, and 
it is not present, instead, at RHIC energy ($\sqrtsNN=200$ GeV, STAR)~\cite{StarSQM07}. 
\newline
{\bf Energy dependence}
We measured hyperon production in pBe and Pb--Pb collisions at 40 A GeV/$c$\ allowing 
us to establish that the enhancement mechanism is already effective at this energy, 
with the same 
hierarchy with the strangeness content 
as that observed at higher energy.   
For the most central collisions the
enhancements are higher at 40 than at 158 GeV/$c$,  
the increase with $N_{wound}$\ is steeper at 40 than at 158 GeV/$c$.   
Similar enhancement pattern (except for the \PagL)
has been measured at $\sqrtsNN=$200 GeV/$c$~\cite{StarSQM07}.  

Both dependencies put constraints on theoretical models: e.g.,  
the {\it ``canonical suppression''} model~\cite{canonical} predicts a higher enhancement  
at 40 GeV, but fails to reproduce the absolute amounts and  
the centrality dependence of the enhancements.  
\section{Strange particle spectra}
\begin{figure}[hbt]
\centering
\resizebox{0.33\textwidth}{!}{%
\includegraphics{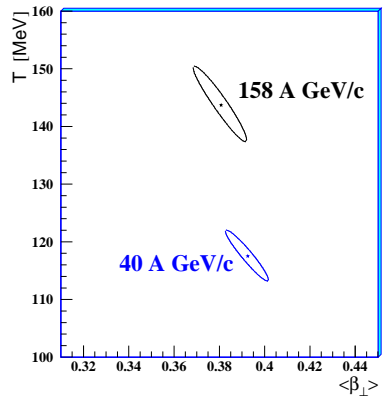}
}
\caption{
Contour
plots in the $T$--$\av{\beta_{\perp}}$ plane at the 1$\sigma$\ 
confidence level. 
}
\label{fig:Cont}
\end{figure}

The bulk of the kinematical distributions of the particles produced in nucleus--nucleus 
interactions are expected to be shaped by the superposition of two effects: the 
thermal motion of the particles in the {\it fireball} and a pressure-driven radial 
(for $\mt=\sqrt{\pt^2 +m^2}$) or longitudinal (for $y$) collective flow, induced 
by the fireball expansion.   
Studies of the $\mt$\ spectra for \PgL, $\Xi$, $\Omega$ hyperons and \PKzS, 
measured in Pb--Pb collisions at 158 and 40 A GeV/$c$\ were presented 
in~\cite{BlastPaper} and~\cite{Blast40}, respectively. A study of the rapidity 
distributions at 158 A GeV/$c$\ was presented in~\cite{rapidity}. 
Here we shall concentrate on  
the $\mt$ spectra, which  
were analyzed in the frame-work of the {\it blast-wave}  
model~\cite{BlastRef}. 
The model assumes cylindrical symmetry for an expanding fireball in local
thermal equilibrium and predicts the shape of the
double-differential yield $\d^2 N/\d \mt \d y$
for the different particle species, in terms of the kinetic freeze-out
temperature $T$ and of
the radial velocity profile $\beta_{\perp}(r)$.   
The latter was parametrized as $\beta_{\perp}(r) = \beta_S\cdot r/R_G$,
$\beta_S$\ being the flow velocity at the freeze-out surface and $R_G$ the
outer radius.
The fit to the experimental spectra allows us to extract $T$ and the average
transverse flow velocity $\av{\beta_{\perp}}=2/3\cdot\beta_S$.
The 1$\sigma$\ contour plot in the freeze-out parameter space is shown 
in figure~\ref{fig:Cont} (left panel) for the most central 53\% of the inelastic 
Pb--Pb  cross section at 40 and 158 A GeV/$c$. More detailed 
analyses of the spectra~\cite{BlastPaper,Blast40} suggested that multi-strange 
hyperons could possibly undergo an earlier freeze-out than singly strange particles.   

\begin{figure}[hbt]
\centering
\resizebox{0.99\textwidth}{!}{%
\includegraphics{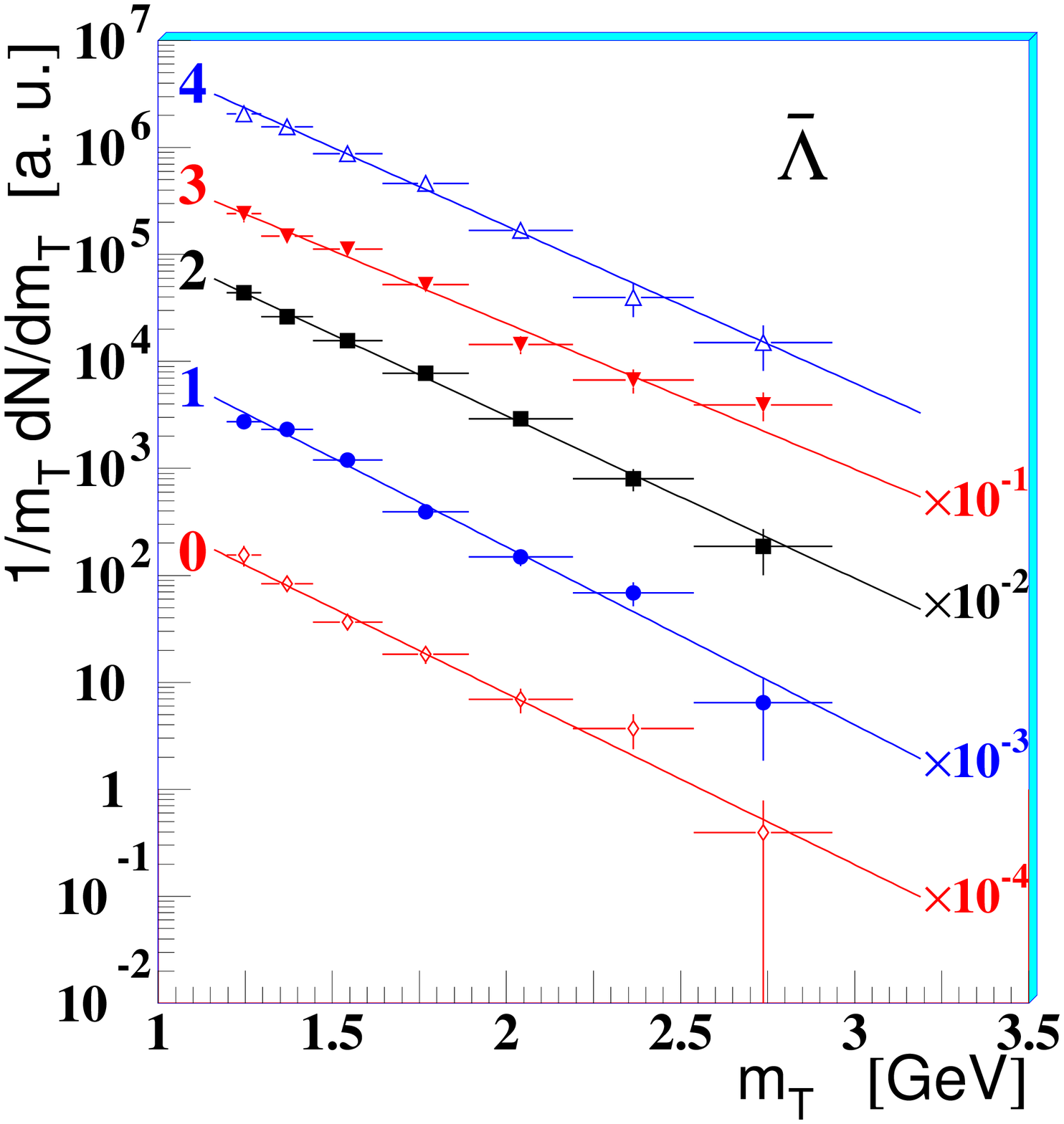}
\includegraphics{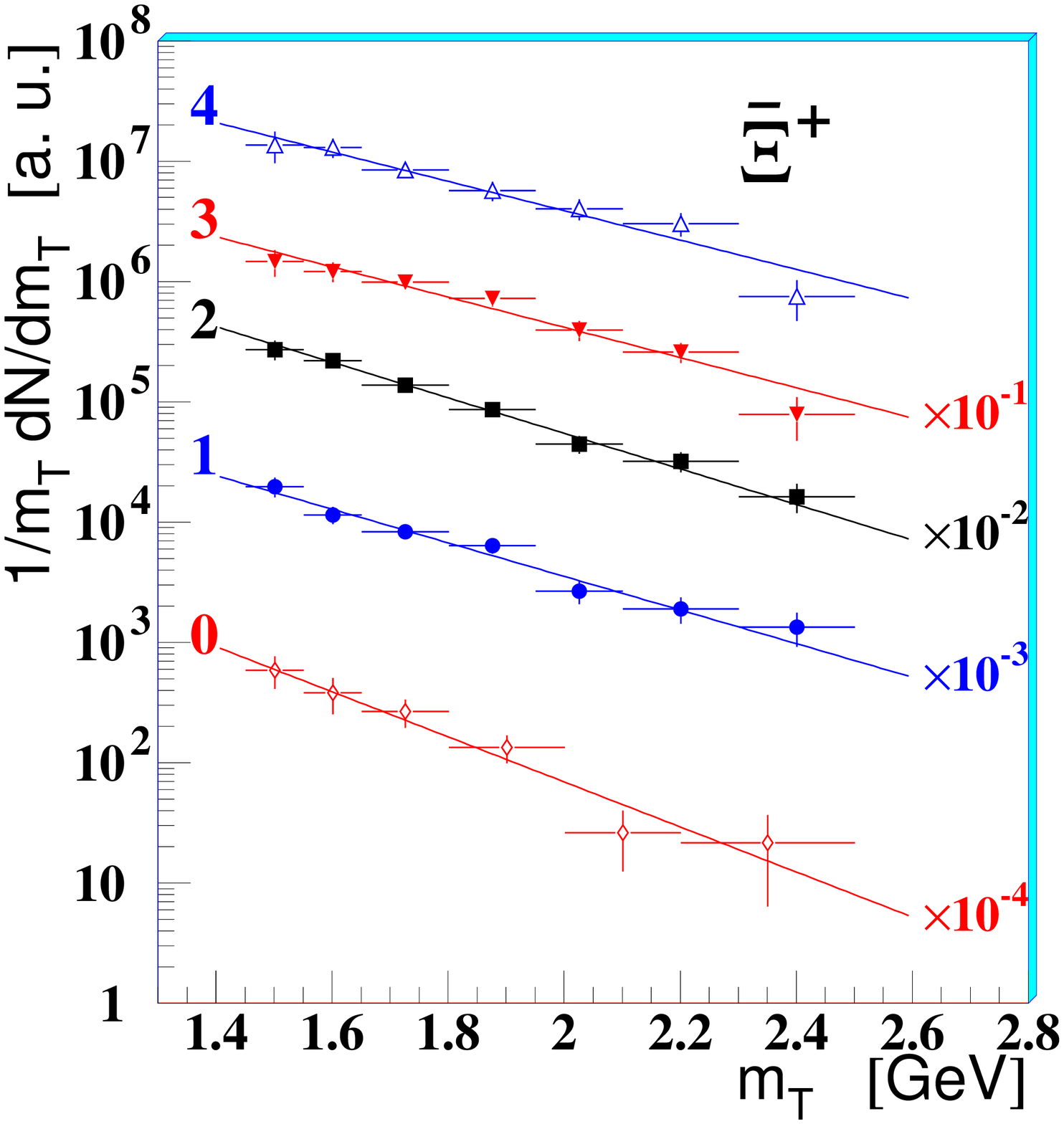}
\includegraphics{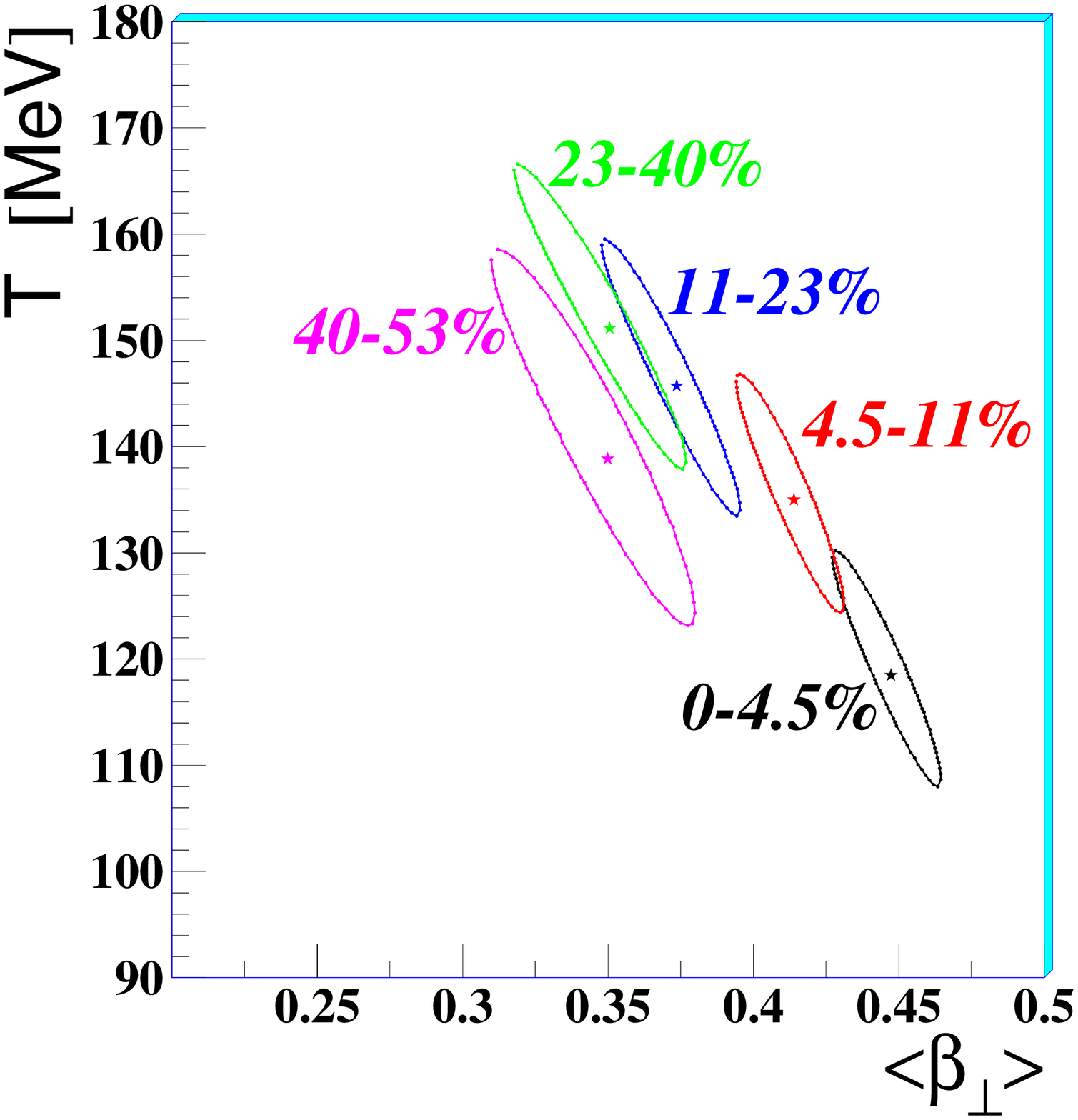}
}
\caption{\rm  Transverse mass spectra for 
              \PagL\ (left panel) and \PagXp\ (middle pannel)        
              from Pb--Pb collisions at 158 A GeV/$c$\ 
              for different centralities. 
              Class $4$, displayed uppermost,  corresponds to the most central collisions (5\%); 
              class $0$, displayed lowermost, corresponds to the most peripheral ones(40--53\%).
              Right: 1$\sigma$\ confidence level contours from fits of all strange particle 
              spectra for different centrality classes.}
\label{fig:msd_spectra}
\end{figure}

Similar results 
were reported at the SPS by the NA49 experiment~\cite{NA49Blast}.  
NA57 has measured the $\mt$\ spectra also as a function of centrality,   
as shwon, e.g., in left and middle panels of figure~\ref{fig:msd_spectra}.   
Therefore we could explore the centrality  
dependence  of the freeze-out parameters, which can be summarized as follows: 
the more central the collisions the 
stronger  
the transverse collective flow and 
the lower the final thermal freeze-out temperature. This finding 
is very relevant to  
the question of whether observables at the SPS 
(e.g. the elliptic flow) can    
be described hydro-dynamically, as it is the case at RHIC  
(see~\cite{BrunoHQ} for a discussion).  

\section{Nuclear modification factors $\RCP$}
\begin{figure}[hbt]
\centering
\resizebox{0.99\textwidth}{!}{%
\includegraphics{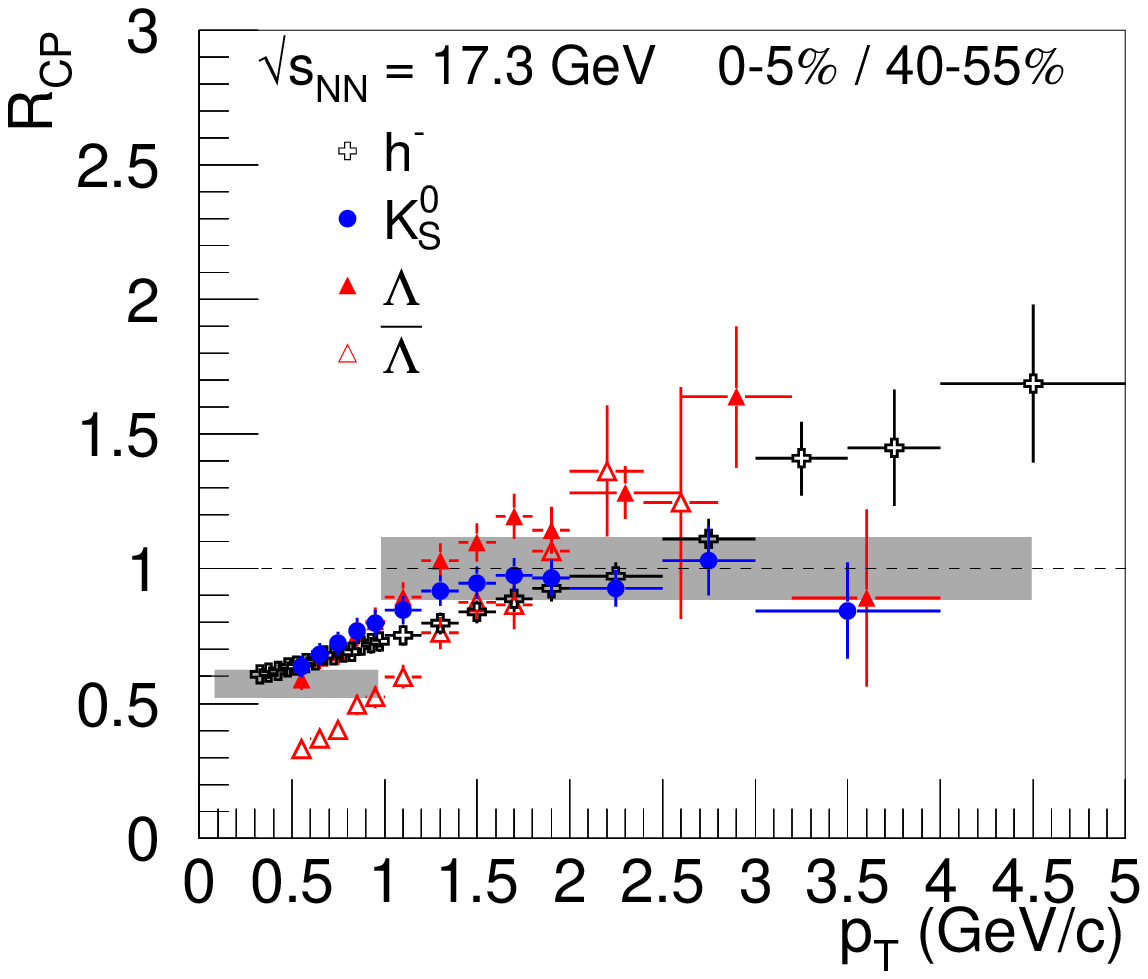}\hspace{2.9cm}
\includegraphics{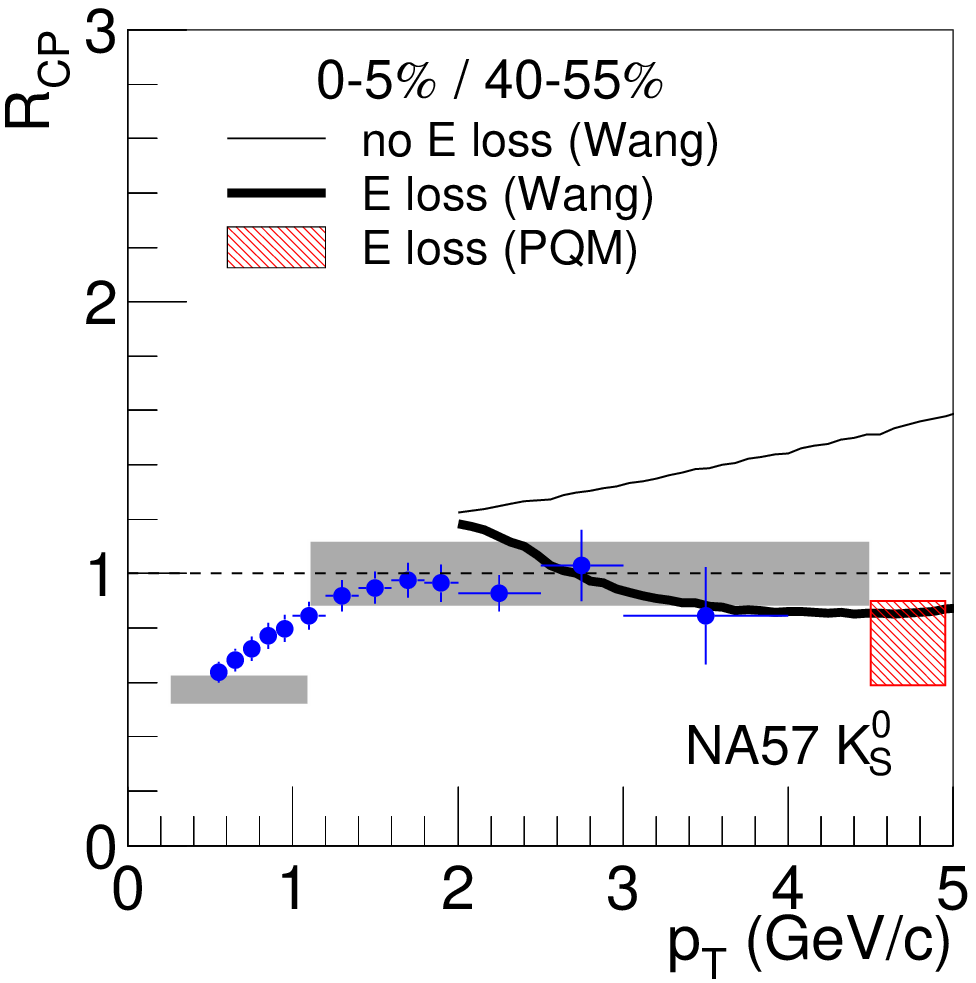}
\includegraphics{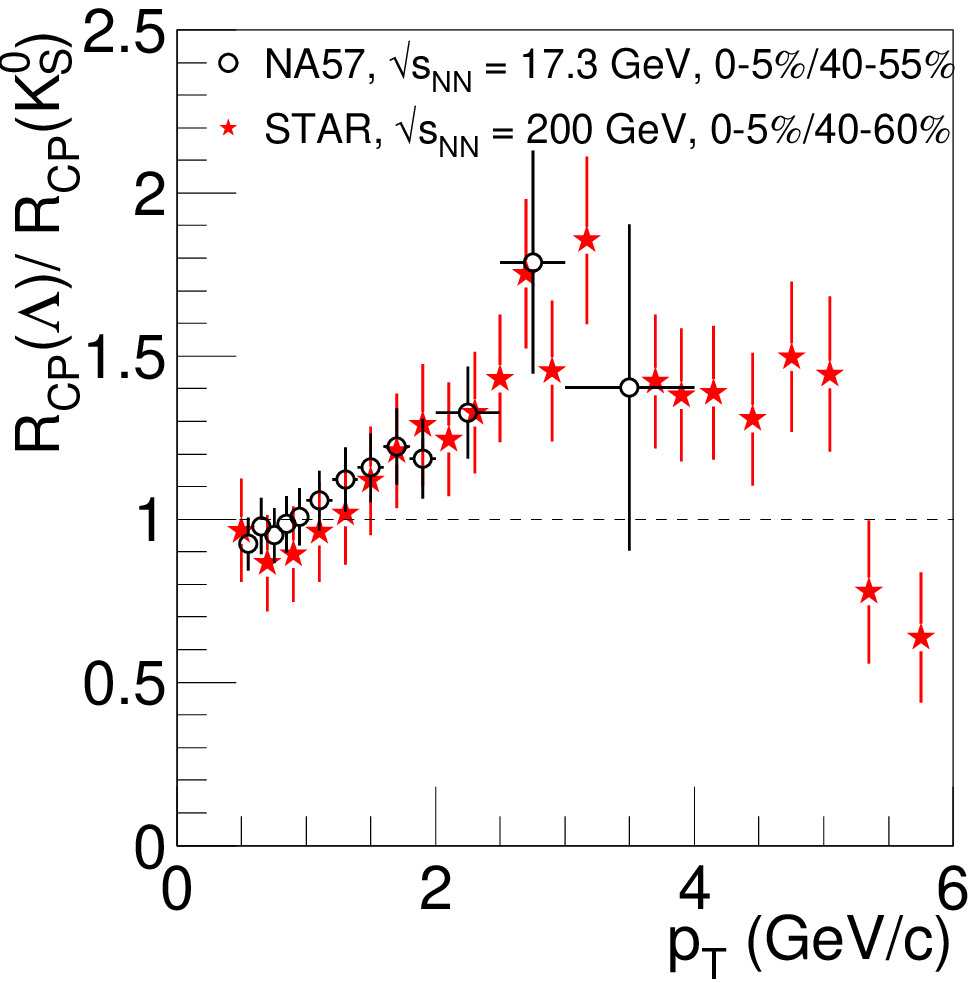}
}
\caption{Left: $\RCP$ ratios for negatively charged particles ($h^-$) and
           singly-strange particles in
           Pb--Pb collisions at $\sqrtsNN=17.3~\gev$~\cite{rcppaper}.
            The width of the shaded band centered
            at $\RCP=1$ indicates the systematic error due to the uncertainty
            on the values of $\av{N_{\rm coll}}$\ in each class; the band
            at low $\pt$ show the value expected for
            scaling with the number of participants.
          Middle:
            $\rm K^0_S$ $\RCP(\pt)$\
            compared to predictions~\cite{wang,pqm}
            with and without energy loss.
          Right: 
            ratio of $\Lambda$ $\RCP$ to $\rm K^0_S$ $\RCP$,
            as a function of $\pt$, at the
            SPS (NA57 at $\sqrtsNN=17.3~\gev$) and at
            RHIC (STAR at $\sqrtsNN=200~\gev$~\cite{starRcpk0la}).
            }
\label{fig:Rcp}
\end{figure}

At the Relativistic Heavy Ion Collider (RHIC),
the central-to-peripheral nuclear modification factor
\[
\RCP(\pt) = {\av{N_{\rm coll}}_{\rm P} \over \av{N_{\rm coll}}_{\rm C}}\times
\frac{{\rm d}^2 N_{\rm AA}^{\rm C}/{\rm d}\pt{\rm d} y}{{\rm d}^2 N_{\rm AA}^{\rm P}/{\rm d}\pt{\rm d} y}
\]
 measured for a large variety of particles
has proved to be a powerful tool for the study of parton propagation
in the dense QCD medium expected to be formed in nucleus--nucleus collisions
(see, e.g.,~\cite{starRcpk0la}). 
At SPS energies, only $\pi^0$\ $\RCP$\ measurements were 
available~\cite{wa98} prior to NA57; we reported the  first results on the
particle species dependence (unidentified 
negatively charged hadrons, \PKzS, \PgL, and \PagL)
in~\cite{rcppaper}.
Figure~\ref{fig:Rcp} (left panel) shows the results for 0--5\%/40--55\%
$\RCP$ nuclear modification factors.
At low-$\pt$ $\RCP$ scales with the number of participants for
all particles except the $\overline\Lambda$. With increasing $\pt$,
$\rm K_S^0$ mesons reach values of $\RCP\approx 1$;   
we did not observe the enhancement above unity that was measured in
proton--nucleus relative to pp collisions 
(Cronin effect~\cite{cronin}). An enhancement is,
instead, observed for strange baryons, $\Lambda$ and $\overline\Lambda$,
that reach $\RCP\simeq 1.5$ at $\pt\simeq 3~\gev/c$.
In fig.~\ref{fig:Rcp} (middle panel)
we compare our $\rm K^0_S$ data with predictions 
obtained from a pQCD-based  
calculation~\cite{wang}, including (thick line) or
excluding (thin line) in-medium parton energy loss. The initial
gluon rapidity density of the medium
was scaled down from that
needed to describe RHIC data, according to the decrease by
about a factor 2 in the charged particle multiplicity.
The data are better described by the curve that includes energy loss.
The prediction of a second model of parton energy loss (PQM) that describes several
energy-loss-related observables at RHIC energies~\cite{pqm} is also
in agreement with the first model.
Figure~\ref{fig:Rcp} (right panel) shows the ratio
of $\Lambda$ $\RCP$ to ${\rm K^0_S}$ $\RCP$, as measured from our data
and by STAR at $\sqrtsNN=200~\gev$~\cite{starRcpk0la}
(note that the sum $\Lambda+\overline\Lambda$ is shown for STAR results).  
The similarity of the $\Lambda$--$\rm K$ pattern
to that observed at RHIC can be taken as an indication for coalescence
effects~\cite{coalescence} at SPS energy.
\section{Conclusions}
The planned experimental programme of NA57 has been succesfully
carried through. The experiment has made several substantial
contributions to the study of the properties of the new state formed in 
heavy-ion collisions at the SPS, particularly in three areas: the study 
of the hyperon enhancements and of their dependence on strangeness
content, collision centrality and collision energy; the study of the 
transverse expansion and freeze-out conditions, and in particular their 
centrality and energy dependence; the study of the nuclear modification 
factors at the SPS. 
%
%

\end{document}